\documentclass{aa}
\usepackage{psfig}
\newcommand{\ltap}{\mathrel{\hbox{\rlap{\lower.55ex \hbox {$\sim$}}
                   \kern-.3em \raise.4ex \hbox{$<$}}}}

\begin{document}

\thesaurus{05(08.01.2; 08.02.3; 10.15.2; 13.25.5)}

\title{Optical spectroscopy of X-ray sources in the old open\\ cluster M$\,$67 
\thanks{Based on observations made with the William Herschel Telescope 
operated on the island of La Palma by the Isaac Newton Group in the 
Spanish Observatorio del Roque de los Muchachos of the Instituto de
Astrofisica de Canarias}}

\titlerunning{Optical spectroscopy of X-ray sources in M$\,$67}

\author{Maureen van den Berg\inst{1} \and Frank Verbunt\inst{1} \and 
Robert D. Mathieu\inst{2}}

\authorrunning{Maureen van den Berg et al.}

\offprints{Maureen van den Berg}

\mail{m.c.vandenberg@astro.uu.nl}

\institute{   Astronomical Institute,
              P.O.Box 80000, 3508 TA Utrecht, The Netherlands
         \and Department of Astronomy, University of Wisconsin,
              Madison, WI~53706, U.S.A.
                }

\date{Received date / Accepted date}   

\maketitle

\begin{abstract}
We have obtained optical spectra of seven stars 
in the old galactic cluster \object{M$\,$67} that are
unusual sources of X-rays, and investigate whether the
X-ray emission is due to magnetic activity or to mass transfer.
The two binaries below the giant branch S\,1063 and S\,1113, the giant with 
the white dwarf companion S\,1040 and the eccentric binary on the subgiant 
branch S\,1242 show magnetic activity in the form of
\ion{Ca}{ii} H\&K emission and H$\alpha$ emission, suggesting that their
X-rays are coronal. The reason for the enhanced
activity level in S\,1040 is not clear. The two wide, eccentric binaries 
S\,1072 and S\,1237 and the blue straggler
S\,1082 do not show evidence for \ion{Ca}{ii} H\&K emission.
A second spectral component is found in the spectrum of S\,1082,
most clearly in the variable H$\alpha$ absorption profile. We interpret this
as a signature of the proposed hot subluminous companion.
\keywords{Stars: activity -- binaries: general -- open clusters and 
associations: individual: M$\,$67 -- X-rays: stars}
\end{abstract}

\section{Introduction}

\begin{figure}
\centerline{\psfig{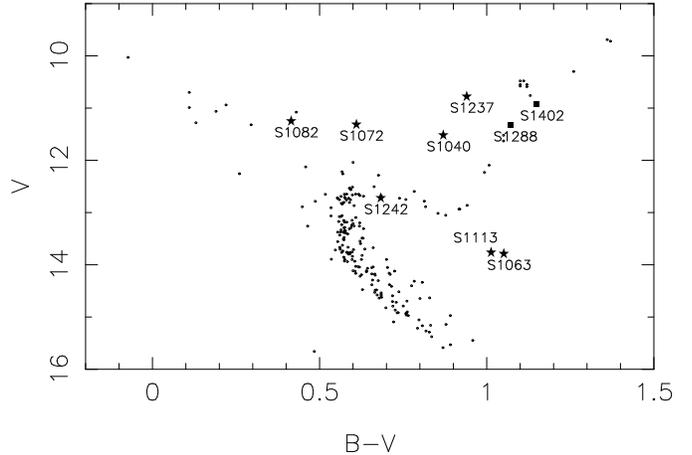} {\hfil}}
\caption{Colour-magnitude diagram of \object{M$\,$67}. Colours and magnitudes 
are from
Montgomery et al.\ (1993). Only stars with membership probability 
$> 0.8$ (based on their proper motion, Sanders 1977) were selected.
Stars indicate the observed X-ray binaries; squares two member giants
observed for comparison.
} \label{cmd}
\end{figure}

Two observations of \object{M\,67} with the ROSAT PSPC resulted 
in the detection of X-ray emission from 25 members of this old open cluster 
(Belloni et al.\ 1993, 1998). 
\nocite{bellea} \nocite{bellea93}
The X-ray emission of many of these sources is readily understood.
For example, the X-ray emission originates in deep, hot atmospheric layers
in a hot white dwarf; is due to mass transfer in a cataclysmic variable;
and is caused by magnetic activity in two contact binaries 
and several RS CVn-type binaries.
However, Belloni et al.\ (1998) point out several X-ray sources in 
\object{M$\,$67}
for which the X-ray emission is unexplained.
All but one of these objects are located away from the isochrone formed by 
the main sequence and the (sub)giant branch of \object{M$\,$67} (Fig.~\ref{cmd}). 

In this paper we investigate the nature of the X-ray emission of these
stars through low- and high-resolution optical spectra.
In particular, we investigate whether the emission could be coronal as
a consequence of magnetic activity, by looking for emission cores in the 
\ion{Ca}{ii} H\&K lines. Tidal interaction in a close binary 
orbit is thought to enhance
magnetic activity at the stellar surface by spinning up the stars in the 
binary. Therefore, we also derive projected rotational velocities with the 
crosscorrelation
method. Finally, we study the H$\alpha$ profile as a possible indicator of 
activity or mass transfer.

The observations and the data reduction are described in Sect.~2,
and the analysis of the spectra in Sect.~3.
Comparison with chromospherically active binaries is made in Sect.~4.
A discussion of our results is given in Sect.~5.
In the remainder of the introduction we give brief sketches of the
stars studied in this paper; details on many of them are given by
Mathieu et al.\ (1990). The stars are indicated with their number in
Sanders (1977), and are listed in Table~\ref{tab1}.
\nocite{mathlathea} \nocite{san}

\begin{table}
\caption{Stars of \object{M$\,$67} discussed in this paper. Visual
magnitude and colour (from Montgomery et al.\ 1993), spectral type
(from Allen \&\ Strom 1995, and Zhilinskii \&\ Frolov 1994), orbital
period and eccentricity (from Mathieu et al. 1990, Latham et al.\ 1992 
and -- for S\,1113 --
Mathieu et al.\ 1999, in preparation), and X-ray countrate in PSPC channels 
41--240,
corresponding to 0.4--2.4\,keV (from Belloni et al.\ 1998).
\label{tab1}}
\begin{tabular}{l@{\hspace{0.25cm}}l@{\hspace{0.25cm}}l@{\hspace{0.25cm}}l@{\hspace{0.12cm}}l@{\hspace{0.25cm}}l@{\hspace{0.25cm}}l}
ID    & $V$   & $B$-$V$ & sp.type & $P_{\mathrm{b}}$ & $e$ & ctrate \\
      &       &       &         &    (d)           &     &   (s$^{-1}$) \\  
 \\
\hline
 \\
S\,1040 & 11.52 & 0.87 & G4III    & 42.83 & 0.027(28)       & 0.0050(6) \\
S\,1063 & 13.79 & 1.05 & G8IV     & 18.39 & 0.217(14)       & 0.0047(6) \\
S\,1072 & 11.32 & 0.61 & G3III--IV & 1495. & 0.32(7)         & 0.0013(3) \\
S\,1082 & 11.25 & 0.42 & F5IV     &       &                 & 0.0046(6) \\
S\,1113 & 13.77 & 1.01 &          & 2.823  & 0.031(14)       & 0.0047(6) \\
S\,1237 & 10.78 & 0.94 & G8III    & 697.8 & 0.105(15)       & 0.0010(3) \\
S\,1242 & 12.72 & 0.68 &          & 31.78 & 0.664(18)       & 0.0007(2) \\
\end{tabular}
\end{table}
\nocite{montea} \nocite{lathmathea} \nocite{allestro} \nocite{zhilfrol}
\nocite{mathlathea}

S\,1063 and S\,1113 are two binaries located  below the subgiant branch
in the colour-magnitude diagram of \object{M$\,$67}.
Their orbital periods, 18.4 and 2.82 days respectively, are too long
for them to be contact binaries; also they are too far above the main
sequence to be binaries of main-sequence stars.
In principle, a (sub)giant can become underluminous when it transfers
mass to its companion, as energy is taken from the stellar luminosity
to restore hydrostatic equilibrium (e.g.\ Kippenhahn \&\ Weigert 1967).
However, mass transfer through Roche lobe overflow very rapidly leads to
circularization of the binary orbit, whereas S\,1063 has an
eccentricity $e=0.217$. 
The orbit of S\,1113 is circular, so mass transfer could be occurring 
in that system.
For the moment, the nature of these binaries is not understood.
In both, Pasquini \&\ Belloni (1998) observed emission cores in the 
\ion{Ca}{ii} H\&K lines. 
S\,1063 is reported to be photometrically variable with $\sim$ 0.10 mag 
(Rajamohan et al.\ 1988; Kaluzny \& Radczynska 1991), 
but no period is found. 
For S\,1113, photometric variability with a period of 0.313 days
and a total amplitude of 0.6 mag was claimed by Kurochkin (1960), but this 
has not been confirmed by Kaluzny \& Radczynska (1991), who find variability
with only 0.05 mag. S\,1063 is the only \object{M\,67} star in our sample that
shows significantly variable X-ray emission (between 0.0081 and 0.0047 
cts s$^{-1}$; Belloni et al.\ 1998). 
\nocite{kippweig} \nocite{rajaea} \nocite{kuro} \nocite{kara} \nocite{pasqbell} 

S\,1072 and S\,1237 are binaries with orbital periods of 1495 and 698
days, and with eccentricities $e=0.32$ and 0.105, respectively.
The colour and magnitude of S\,1072 cannot be explained with the pairing
of a giant and a blue straggler, since this is not compatible with its
$ubvy$ photometry (Nissen et al.\ 1987; Mathieu \&\ Latham 1986), nor with superposition of three
subgiants, since this is excluded by the radial velocity correlations 
(Mathieu et al.\ 1990).
The absence of the 6708 \AA\ lithium 
feature in the spectrum of S\,1072 indicates that the surface material has 
undergone mixing (Hobbs \&\ Mathieu 1991; Pritchett \& Glaspey 1991).
S\,1237 could be a binary of a giant and a star at the top of the
evolved main sequence (Janes \& Smith 1984); high-resolution spectroscopy should
be able to detect the main-sequence star in that case (Mathieu et al.\ 1990).
The wide orbits and significant eccentricities appear to exclude
both mass transfer and tidal interaction as explanations for the X-ray
emission.
\nocite{nissea} \nocite{hobbmath} \nocite{pritglas} \nocite{janesmit}
\nocite{mathlath}

S\,1242 has the largest eccentricity of the binaries in our sample,
at $e=0.66$ in an orbit of 31.8$\,$days. Its position on the subgiant
branch is explained if a subgiant of 1.25\,$M_{\sun}$ has a secondary with $V>15$
(Mathieu et al.\ 1990). \ion{Ca}{ii} K line emission is 
reported by Pasquini \&\ Belloni (1998).
Photometric variability with a period of 4.88$\,$days and amplitude of 0.0025
mag has been found by Gilliland et al.\ (1991).
We note that this photometric period corresponds to corotation
with the orbit at periastron, which suggests that the X-ray emission
may be due to tidal interaction taking place at periastron.
The binary would then be an interesting example of a system in transition 
from an eccentric to a circular orbit. Indeed, according to the diagnostic
diagram of Verbunt \&\ Phinney (1995) a giant of 1.25\,$M_{\sun}$ with a
current radius of $\simeq2.3\,R_{\sun}$ (as derived from the location of 
S\,1242 in the colour-magnitude diagram) cannot have circularized an orbit
of 31.8\,days.
\nocite{gillea} \nocite{verbphin}

S\,1040 is a binary consisting of a giant and a white dwarf. The progenitor
of the white dwarf circularized the orbit during a phase of
mass transfer (Verbunt \&\ Phinney 1995); as a result the mass of the
white dwarf is very low (Landsman et al.\ 1997).
The white dwarf is probably too cool, at 16\,160$\,$K, to be the X-ray emitter.
Indications for magnetic activity are \ion{Ca}{ii} H\&K 
(Pasquini \&\ Belloni 1998) and \ion{Mg}{ii} ($\lambda\lambda$ 2800 \AA,
Landsman et al.\ 1997) emission lines.  
If the X-rays are due to coronal emission of the giant, this must be the
consequence of the past evolution of the binary, since the giant is too small 
for significant tidal interaction to be taking place in the current orbit.
\nocite{landea}

S\,1082 is a blue straggler. Photometric variability of 
0.08 mag within a few hours was observed by Simoda (1991). Goranskii 
et al.\ (1992) found eclipses
with a total amplitude of 0.12 mag and a binary period of 
1.07 days; however, the radial velocities of the star
do not show this period, and vary by about 2\,km s$^{-1}$, far too little
for a 1\,day eclipsing binary (Mathieu et al.\ 1986).
Landsman et al.\ (1998) detect a significant excess at 1520 \AA\ with the
Ultraviolet Imaging Telescope, and ascribe this to a hot, subluminous
secondary. Such a secondary was suggested already by Mathys (1991) on the
basis of a broad component in the \ion{Na}{i} D and 
\ion{O}{i} absorption lines.
\nocite{goraea} \nocite{mathlath} \nocite{math} \nocite{landea} \nocite{simo}
\nocite{mathea86}

\nocite{montea} \nocite{san}

\section{Observations and data reduction}

Optical spectra were obtained on February 28/29, 1996 with the 
4.2m William Herschel Telescope on La Palma, under good
weather conditions (seeing $<1\arcsec$ until 4$\fh$30 UT, 
$<2\arcsec$ thereafter). In addition to the X-ray sources in \object{M$\,$67} we 
observed 
two ordinary member giants of \object{M$\,$67}, S\,1288 and S\,1402, for comparison. 
Furthermore one flux standard and three velocity standards
were observed. The blue high-resolution spectra of S\,1113 were obtained on 
April 7/8, 1998 
with the same telescope through a service observation (seeing 1--2$\arcsec$). 
A log of the observations is given in Table~\ref{tab11}.

\begin{table}
\caption{Log of the observations.
For each target we give the UT at start of the exposures and the exposure
time for the ISIS and UES blue and red spectra. All observations were
obtained on 28/29 February 1996, except the blue UES exposures of S\,1113 
which were taken on 7/8 April 1998 (indicated with $^*$ in the table). 
\label{tab11}}
\begin{tabular}{llrlrlr}
ID & \multicolumn{2}{l}{ISIS} & 
\multicolumn{2}{l}{UES blue} & \multicolumn{2}{l}{UES red} \\
 & \multicolumn{2}{l}{---------------} & \multicolumn{2}{l}{---------------}  & \multicolumn{2}{l}{---------------} \\
 & UT & \multicolumn{1}{l}{$t_{\mathrm{exp}}$}& UT & \multicolumn{1}{l}{$t_{\mathrm{exp}}$} & UT & \multicolumn{1}{l}{$t_{\mathrm{exp}}$} \\
 & & \multicolumn{1}{l}{(s)} & & \multicolumn{1}{l}{(s)} & & \multicolumn{1}{l}{(s)} \\
 \\ 
\hline \\
\multicolumn{5}{l}{{\sc targets}} \\ 
S\,1040 & 02:43&60 &  22:03&600 & 00:46&300 \\ 
S\,1063 & 02:30&180 & 20:51&1200 & 00:11&1200 \\
              & & & 21:15&1200 & 00:33&600 \\
              & & & 21:37&1200\\ 
S\,1072 & 02:49&60 & 22:24&360 & 01:01&240  \\ 
S\,1082 & 02:39&60 & 20:39&300 & 00:01&360  \\
              & & & 23:24&600 & 01:41&360  \\
                     & & & & &  04:11&360  \\ 
S\,1113 & 03:01&180 & 22:37$^*$ & 900   & 03:36&1200  \\
               & & & 22:58$^*$ & 900  & 03:58&600  \\
S\,1237 & 02:46&60 & 22:16&300 & 00:54&180  \\
S\,1242 & 02:55&120 & 22:33&1500 & 01:08&900 \\ 
\\
\multicolumn{5}{l}{{\sc comparison giants}} \\ 
S\,1288 & 02:58&60 & 23:12&600 & 01:33&300 \\
S\,1402 & 02:52&60 & 23:01&450 & 01:26&240 \\
\\
\multicolumn{5}{l}{{\sc flux standard}} \\ 
 
\object{HZ\,44} & 02:21&80 & 23:42&600 & 01:57&600 \\
       & 03:13&80  \\
       & 04:43&480 \\ 
\\
\multicolumn{5}{l}{{\sc velocity standards}} \\
\object{HD$\,$132737} & & &  06:04& 90 & 05:13&45  \\  
\object{HD$\,$136202} & & &  05:55&100 & 05:42&50  \\ 
\object{HD$\,$171232} & & &  &         & 05:34&45  \\ 
\end{tabular}
\end{table}

All spectra have been reduced using the Image Reduction and Analysis Facility
(IRAF)\footnote{IRAF is distributed by the National Optical Astronomy 
Observatories, which are operated by the Association of Universities for 
Research in Astronomy, Inc., under cooperative agreement with the National
Science Foundation}. 

\subsection{Low-resolution spectra}

Low-resolution spectra were taken with the ISIS double-beam
spectrograph (Carter et al.\ 1993). 
The blue arm of ISIS was used with the 300
lines per mm grating and TEK-CCD, resulting in a wavelength coverage
of 3831 to 5404 \AA\ and a dispersion of 1.54 \AA\ per pixel at
4000 \AA. The red arm, combined with the 316 lines per mm grating and
EEV-CCD, covered a wavelength region of 5619 to 7135 \AA\ with a
dispersion of 1.40 \AA\ per pixel at 6500 \AA. 
The format of the frames is 1124 $\times$ 200 pixels which
includes the under- and overscan regions. For the object exposures the slit
width was set to $4\arcsec$. Flatfields were made with a Tungsten 
lamp while CuAr and CuNe lamp exposures were taken for the purpose of 
wavelength calibration.
\nocite{cartea}

For the ISIS-spectra, basic reduction steps have been
done within the IRAF {\sc ccdred}-package. These steps include removing the 
bias signal making use of the under- and overscan regions and zero frames, 
trimming the 
frames to remove the under- and overscan, and flatfielding to correct for small
pixel-to-pixel gain variations. The remaining reduction has been done with 
IRAF {\sc specred}-package tasks. 
With the optimal extraction algorithm (Horne 1986)
the two dimensional images are reduced to one dimensional spectra. 
Next, the spectra are calibrated in wavelength with the arc frames. A 
dispersion solution is found by fitting third (blue) and fourth (red) order 
polynomials to the positions on the CCD of the arclamp lines. 
The fluxes of the spectra are calibrated with the absolute fluxes of HZ\,44, 
tabulated at 50 \AA\ intervals (Massey et al. 1988), and adopting the
standard  atmospheric extinction curve for La Palma as given by King (1985).
The estimated accuracy of the flux calibration is $\sim 10$\%.
\nocite{horn}\nocite{king} \nocite{massea} 

\subsection{High-resolution spectra}

High-resolution echelle spectra were taken with~the Utrecht Echelle 
Spectrograph (UES, Unger et al.\ 1993). 
Observations were done with a $\sim 1\arcsec$ slitwidth. 
\nocite{ungeea} 

For the 1996 observations, the UES was used in combination with a 
1024 $\times$ 1024 pixels TEK-CCD,
and the 31.6 lines per mm grating (E31), which resulted in a broad wavelength 
coverage, but small separation of the echelle orders on the CCD. 
In this setup, the UES resolving power is 49\,000 per resolution element 
(two pixels), corresponding to a dispersion of 3 km s$^{-1}$ per pixel
or 0.06 \AA\ per pixel at 6000 \AA. 
The frames were centered on $\lambda_\mathrm{cen}=4250$ \AA\ and 
$\lambda_\mathrm{cen}=5930$ \AA\ in order to get a blue (3820 to 4920 \AA) and
red (4890 to 7940 \AA) echelle spectrum. The number of orders recorded on the 
CCD is 34 in the blue and 45 in the red, each covering $\sim$ 45 to 80 \AA\
increasing for longer wavelengths. Towards the red, gaps occur between the
wavelength coverage of adjacent orders. 
Exposures of a quartz lamp were taken to make the flatfield corrections.
ThAr exposures served as wavelength calibration frames.

For the 1998 observations of S\,1113, a 2048 $\times$ 2048 pixels SITe-CCD was 
used. Two spectra were taken with the 79.0 lines per mm (E79) grating 
($\lambda_\mathrm{cen}=4343$ \AA). The difference between the E79 and the E31
gratings is that E79-spectra have a larger separation of the echelle-orders 
on the detector, which can improve the determination of the sky-background. 
The spectral 
resolution of the gratings is the same (the central dispersion in these
observations is $\sim 0.04$ \AA\ per pixel). The specified wavelength-coverage 
for this combination of grating and detector is 3546 to 6163 \AA\, but only the 
central orders were bright enough to extract spectra (3724 to 5998 \AA). 
Flatfield and ThAr
exposures were made for calibration purposes. 

The reduction of the UES spectra has been performed using the routines 
available within the IRAF {\sc echelle}-package. 
First, the frames are debiased and the under- and overscan regions removed. 
After locating the orders on the CCD for both the quartz lamp and the object 
exposures, we flatfielded the frames.
Spectra are extracted with optimal extraction. 
The small order separation makes sky subtraction difficult; however, our
targets are bright, and the resulting error is negligible.
In the step of wavelength
calibration, the dispersion solution is derived by fitting
third and fourth order polynomials leaving rms-residuals of 0.004 \AA\ (red)
and 0.002 \AA\ (blue, 0.003 \AA\ for the 1998-spectra). To find absolute fluxes for the \ion{Ca}{ii} K 
($\lambda \, 3933.67$ \AA) \& H ($\lambda \, 3968.47$ \AA) emission lines 
(Sect.~3.1), the fluxes of the relevant blue orders of an object have been 
calibrated with the calibrated ISIS spectrum of the same object. Continuum 
normalization of the orders in the red spectra, required for the rotational 
velocity analysis, is done by fitting third to fifth order polynomials to the 
wavelength-calibrated spectra.

\section{Data analysis}

We study two indicators of magnetic activity.
The direct indicator is emission in the cores of the
\ion{Ca}{ii} H\&K lines.
Another indicator is the rotational speed: rapid (differential) rotation 
and convective motions are thought to generate magnetic fields through a 
dynamo.

\subsection{Determination of \ion{Ca}{ii} H\&K emission fluxes}

To estimate the amount of flux emitted in the \ion{Ca}{ii} H\&K line
cores, $F_{\mathrm{Ca}}$, we add the fluxes above the H\&K absorption 
profiles as follows. An upper and a lower limit of the
level of the absorption pseudo-continuum is estimated by eye and 
is marked by a straight line. For S\,1113 this is illustrated in Fig.~\ref{subca}.
We obtain a lower and upper limit 
of the emitted flux by adding the fluxes in each wavelength-bin 
above these levels. The value given in Table~\ref{fluxes} is the 
average of these two results, the uncertainty is half their difference. 
Use of higher order fits (following Fern\'andez-Figueroa et al.\ 1994)
to the absorption profile gives similar results.
If an emission line is not clearly visible, we obtain an upper limit by 
estimating the minimal detectable emission flux at the H\&K line centers
within a 1 \AA\ wide region (typical width of the emission lines). 

\begin{figure}
\centerline{\psfig{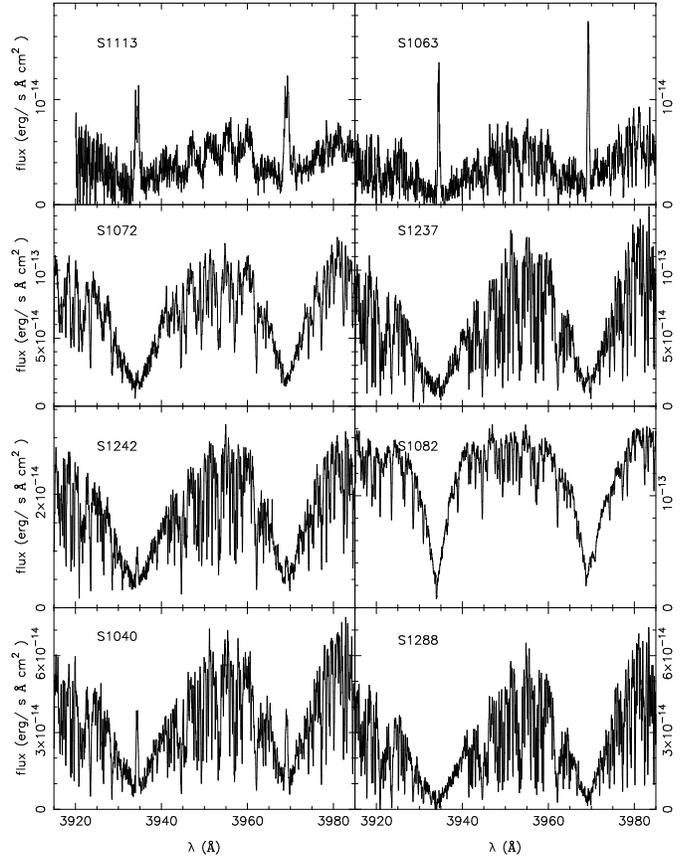} {\hfil}}
\caption{The \ion{Ca}{ii} H\&K regions in the high-resolution spectra of 
our targets. In the lower right corner the spectrum of the non-active 
comparison giant S\,1288 is shown.} \label{cahk}
\end{figure}

Six of our sources show \ion{Ca}{ii} H\&K line emission (Fig.~\ref{cahk}).
The profiles of S\,1113 appear to be double-lined, suggesting that we see activity 
of both stars (Fig.~\ref{subca}). The fluxes given in Table~\ref{fluxes} are the total fluxes, 
i.e.\ no attempt was made to deblend the emission lines. 
No emission is visible in the spectrum of S\,1082.

\begin{table}
\caption{Fluxes of emission cores in \ion{Ca}{ii} H\&K lines and
projected rotational velocities derived from our high-resolution spectra.
The velocities were determined from crosscorrelation with the spectrum 
of \object{HD\,136202} (F8III-IV, ESA 1997) for S\,1040, S\,1072, S\,1082 and S\,1242; and with
the spectrum of \object{HD\,171232} (G8III, ESA 1997) for the other stars. Both HD stars have
line widths $\tau=1.8\pm$0.1\,km s$^{-1}$. For S\,1113 we list both components; the primary contributes $\sim 82$\%\ of
the light and has broader lines.
\label{fluxes}}
\begin{tabular}{lrrl}
ID & \multicolumn{1}{l}{$\log F_{\ion{Ca}{k}}$} & \multicolumn{1}{l}{$\log F_{\ion{Ca}{h}}$} & \multicolumn{1}{l}{$v\sin i$} \\
   & \multicolumn{1}{l}{(${\rm erg}\,{\rm cm}^{-2}\,{\rm s}^{-1}$)} & \multicolumn{1}{l}{(${\rm erg}\,{\rm cm}^{-2}\,{\rm s}^{-1}$)}  & 
\multicolumn{1}{l}{(km s$^{-1}$)} \\ \\ \hline
 \\
 S\,1040 & $16(4)\times10^{-15}$  & $17(5)\times10^{-15}$ & 3.0(1.0) \\
 S\,1063 & $6(1)\times10^{-15}$   & $7(1)\times10^{-15}$  & 3.9(0.8) \\
 S\,1072 & $3(2)\times10^{-15}$   & $<13\times10^{-15}$   & 8.1(1.1) \\
 S\,1082 & $ <11\times10^{-15}$   & $<24\times10^{-15}$   & 5.1(0.7) \\
 S\,1113 & $9(2)\times10^{-15}$   & $10(3)\times10^{-15}$ & 45(6) \\
       &                        &                       & 12(1) \\
 S\,1237 & $4(2)\times10^{-15}$   & $8(5)\times10^{-15}$  & $<1.8$ \\
 S\,1242 & $2.2(0.5)\times10^{-15}$ & $2(1)\times10^{-15}$  & $<2.6(1.0)$ \\
\end{tabular} 
\end{table}
\nocite{hipp}

\subsection{Determination of projected rotational velocities}

\subsubsection{Crosscorrelation}
In order to derive the projected rotational velocity $v\sin i$ of our targets, 
we apply the crosscorrelation technique (e.g. Tonry \& Davis 1979). This 
method computes the correlation between the object spectrum and an 
appropriately chosen template spectrum as function of relative shift. The
position of the maximum of the crosscorrelation function ({\sc ccf}) provides 
the value of the radial-velocity difference between object and template. The 
width of the peak is indicative for the width of the spectral lines and can 
therefore be used as a measure of the rotational velocity of the stars.
\nocite{tonrdavi} 

For rotational velocities not too large the line profiles may be
approximated with Gaussians, allowing analytical treatment 
of the crosscorrelation method.
Assuming that the binary spectrum is a shifted, scaled and 
broadened version of the template spectrum, the broadening 
can be related to the width of the {\sc ccf} peak as follows
(Gunn et al.\ 1996).
With $\tau$ the dispersion in the template's and $\beta$ the 
dispersion in the target's spectral lines,
$\mu$ the dispersion of the {\sc ccf} peak and $\sigma$ the dispersion of the 
gaussian that describes the broadening of the object's spectrum with respect 
to the template, one can write:
\begin{equation}
\mu^2=\tau^2 + \beta^2=2\tau^2 + \sigma^2 
\label{sigma}\end{equation}
Eq.~\ref{sigma} applies to both components in the binary 
spectrum and their corresponding {\sc ccf} peaks. 
\nocite{gunnea}

The crosscorrelations are performed with the IRAF task {\sc fxcor} that uses 
Fourier transforms of the spectra to compute the {\sc ccf}. Before performing 
the crosscorrelation, the continuum is subtracted from the normalized spectra. 
Filtering in the Fourier domain is applied to avoid undesirable contributions 
originating from noise or intrinsically broad lines (see Wyatt 1985). 
\nocite{wyat}  

The templates are chosen from the radial-velocity standards such that their 
spectral types resemble those of the targets. 
The value of $\tau$ for these stars is determined for each order separately by 
autocorrelation of the template spectrum adopting the same filter used for
the crosscorrelations.
In this case $\sigma$ is zero and therefore $\tau$ is found directly from the 
width of the {\sc ccf} peak: ${\tau}^2={\mu}^2/2$.

The template spectra are correlated with our target stars
order by order, where we limit ourselves to orders in the red spectra that do 
not suffer from strong telluric lines.  Most {\sc ccf} peaks can be 
fitted well with a gaussian. As the final value for $v\sin i$ we give
the broadening $\sigma$ averaged over the different orders, and for 
the uncertainty we take the rms of the  spread around the average of $\sigma$
(Table~\ref{fluxes}).
Equating $v\sin i$ to $\sigma$ implicitly assumes that $\tau$ is
the width of the lines not related to rotation. An upper limit
to $v\sin i$ is found from the other extreme in which we assume that the total
width of the spectral line $\beta$ follows from
rotation. This creates uncertainties of the order of $\sigma$ for 
S\,1242. In the case of S\,1237 we find that $\beta < \tau$ ($\sigma < 0$), 
i.e. the lines in S\,1237 are narrower than in the template. For these two
stars, we give an upper limit of $v\sin i < \beta$ in Table~\ref{fluxes}.

S1113 is the only binary observed whose {\sc ccf} shows two peaks.
The {\sc ccf} peaks of both stars overlap in the 1998 observation. 
Therefore we do not use these spectra in the following analysis.
For the 1996 spectra, the {\sc ccf} shows two peaks, one of which is broad 
indicating the presence of a 
fast rotating star. Both peaks are clearly separated with a center-to-center 
velocity separation of $\sim$ 110 km s$^{-1}$. 
The lines in the spectrum are 
smeared out and less pronounced, resulting in a noisy {\sc ccf}. To improve 
this, we combine four sequential orders before 
crosscorrelating (being constrained by the maximum number of points 
{\sc fxcor} can handle).
From eq.~17 in Gunn et al. we derive the relative light 
contribution of both components to the spectrum from the height and 
dispersion of the crosscorrelation peaks, assuming that the binary stars have 
the same spectrum as the template ($\alpha=1$ in eq.~17). According to this
the rapidly rotating star contributes $\sim 82$\%\ of the light.
Note that luminosity ratios derived from cross-correlations are uncertain
and should be confirmed photometrically.

\subsubsection{Fourier-Bessel transformation}
The line profile of the fast rotating star in S\,1113 is not compatible 
anymore with a Gaussian, 
therefore we adopt another method to determine its $v\sin i$, 
 described in Piters et al. (1996). This method uses the 
property that the Fourier-Bessel transform of a spectral line that is purely 
rotationally broadened has a maximum at the position of the 
projected rotational velocity. In practice, this position is a function
of the limits over which the Fourier transform is performed. The local
maxima of this velocity-versus-cutoff-frequency (vcf) function approach 
$v\sin i$ within half a percent, the result growing more accurate for maxima 
at higher frequencies. This error is negligible when compared to errors 
arising from noise, other line broadening mechanisms, etc. (see Piters et al. 
1996). In our determination of $v\sin i$ of the primary of S\,1113, we
have used the first local maximum in the vcf-plot of the transformation of 
four isolated \ion{Fe}{i} lines at $\lambda\lambda$ 6265.14, 6400.15, 6408.03 
and 6411.54 \AA. 
The $v\sin i$ in Table~\ref{fluxes} is an average of the resulting values; 
the $v\sin i$ of the secondary is found from crosscorrelation.

The application of the Fourier-Bessel transformation method is limited
on the low velocity side by the spectral resolution: the Fourier transform
cannot be performed beyond the Nyquist frequency which for slow rotators 
lies at a frequency that is lower than the cutoff-frequency at which the 
first maximum occurs in the vcf-plot. For our spectra this means that
the method cannot be used for $v\sin i <$ 5.2  km s$^{-1}$. Indeed, for every 
star for which the crosscorrelation method gives a $v\sin i$ smaller than 
this value, the vcf-plot does not reach the first local maximum, except
for S\,1082 for which we find $v\sin i$ = 9.5(1.6). For S\,1072, 
the Fourier-Bessel transform gives a $v\sin i$ of 12.7(1.0) km s$^{-1}$. 
Spectral lines were selected from those used in Groot et al. (1996) 
and from the additional lines used for S\,1113.
\nocite{pitegrooea} \nocite{grootea}

\section{Results}

The results of our search for emission cores in the \ion{Ca}{ii} H\&K lines
are displayed in Fig.~\ref{cahk} and in Table~\ref{fluxes}.
The emission lines are strong in S\,1063, S\,1113 and S\,1040;
and still detectable in S\,1242 which indicates chromospheric activity in
these stars.
In S\,1072 and S\,1237 the emission cores are marginal, and on
S\,1082 we can only determine an upper limit.

The (projected) rotational velocities of all of our stars are relatively
small $v\sin i<10\,$km s$^{-1}$, with the exception of S\,1113.

In Sect.~4.1 we investigate whether the relations between X-ray emission,
strength of the emission cores in the \ion{Ca}{ii} H\&K lines, and the
rotational velocities of the unusual X-ray emitters in \object{M$\,$67} are similar
to the relations found for well-known magnetically active stars,
the RS CVn binaries.
In Sect.~4.2 we briefly discuss the behaviour of the H$\alpha$ line
and spectral lines other than \ion{Ca}{ii} H\&K that are indicators of 
chromospheric activity.
Individual systems are discussed in Sect.~4.3.

\subsection{Comparison with RS CVn binaries}

To investigate whether the X-rays of the \object{M$\,$67} stars studied in
this paper are related to magnetic activity, we compare their
optical activity indicators and X-ray fluxes with those of a
sample of RS CVn binaries. In particular, we select RS CVn binaries for which 
fluxes of the emission cores in the \ion{Ca}{ii} H\&K lines have been 
determined from high-resolution spectra by Fern\'andez-Figueroa et al.\ (1994).
To obtain X-ray countrates for these binaries, we searched 
the ROSAT data archive for PSPC observations of them.
We then analyzed all these observations, and determined the countrates, 
in the same bandpass as used in the analysis of \object{M\,67}, 
using the standard procedure described in 
Zimmermann et al.\ (1994).
All pointings that we have analyzed actually led to a positive 
detection of the RS CVn system: even when not the target of the observation,
the RS CVn system is usually the brightest object in the field of view. 
The results of our analysis are listed in Table~\ref{tabros}.
\nocite{fernea} \nocite{zimmea94}

\begin{table}
\caption{ROSAT PSPC countrates for RS~CVn binaries, from our analysis of
archival data. For each source we list the distance, adopted from the Hipparcos
catalogue (ESA 1997), the Julian date ($-$2440000) of the beginning of the
ROSAT exposure, the effective exposure time, the countrate in
channels 41--240 (i.e.\ roughly in the 0.4--2.4\,keV band), and the
offset of the star to the ROSAT pointing direction.
Where applicable we also refer to earlier publication of the ROSAT
observation: $^a$Singh et al. (1996a), $^b$Welty \& Ramsey (1995), 
$^c$White et al. (1994), $^d$Yi et al. (1997), $^e$Singh et al. (1996b), 
$^f$Bauer \&\ Bregman (1996), $^g$Ortolani et al. (1997).
\label{tabros}}
\begin{tabular}{lllrl@{\hspace{0.05cm}}r}
name  &        d    &        JD &  \multicolumn{1}{l}{$t_{\mathrm{exp}}$} & ctrate  & \multicolumn{1}{l}{$\Delta$} \\
      & (pc)    &           & \multicolumn{1}{l}{(s)}                & 
(s$^{-1}$)  & \multicolumn{1}{l}{($\arcmin$)}  \\
 \\ \hline
\\
\multicolumn{6}{l}{{\sc Luminosity class V (Group 1)}} \\ 
\object{IL Com}   &  107$\pm$12  &   8420.575  &   16156      & 0.322(5)   & 36 \\
         &              &   8775.123  &    8345      & 0.126(5)   & 36 \\
\object{TZ CrB}   & 21.7$\pm$0.5 &   8864.531  &    4267      & 6.072(15)  &  0 \\
         &              &   9003.872  &    5776      & 6.228(11)  &  0 \\
         &              &   8864.595  &    3651      & 6.84(2)    & 24 \\
         &              &   9004.871  &    3151      & 5.41(2)    & 24 \\
\object{V772 Her} & 37.7$\pm$1.9 &   9049.117  &   14531      & 1.206(4)   &  0 \\
\object{BY Dra}   & 16.4$\pm$0.2 &   9247.809  &    9895      & 1.009(7)   &  0 \\
\object{V775 Her}$^a$  & 21.4$\pm$0.5 &   9085.714  &    2140      & 1.10(2)    &  0 \\
\object{ER Vul}   & 49.9$\pm$2.1 &   9148.099  &    1209      & 1.23(3)    &  0 \\
\object{KZ And}   & 25.3$\pm$4.9 &   8604.712  &    5249      & 0.888(16)  & 36 \\
\\
\multicolumn{6}{l}{{\sc Luminosity class IV (Group 2)}} \\
\object{V711 Tau} & 29.0$\pm$7   &   8648.446  &    3098      & 6.29(2)    &  0 \\
\object{UX Com}   & 168$\pm$51   &   8426.486  &   21428      & 0.146(3)   & 33 \\
\object{RS CVn}   & 108$\pm$12   &   8810.279  &    2526      & 0.336(17)  & 44 \\
         &              &   8991.040  &    6214      & 0.372(11)    & 44 \\
         &              &   8796.874  &    5076      & 0.305(12)  & 51 \\
         &              &   8966.506  &    2904      & 0.366(18)  & 51 \\
\object{HR 5110}  & 44.5$\pm$1.2 &   8431.422  &   71803      & 2.3152(14) & 44 \\
         &              &   9158.435  &   37658      & 2.403(3)   & 44 \\
\object{SS Boo}$^b$ & 202$\pm$57 &   9030.359  &   10327      & 0.059(2)   &  0 \\
\object{RT Lac}$^b$ & 193$\pm$39 &   8789.392  &    8668      & 0.198(5)   &  0 \\
\object{AR Lac}$^c$ & 42.0$\pm$1.0 &   8620.767  &   13460      & 1.717(4)   &  0 \\
         &              &   9136.873  &    4892      & 2.902(13)  &  0 \\
\\
\multicolumn{6}{l}{{\sc Luminosity class III/II (Group 3)}} \\
\object{12 Cam}   & 192$\pm$34   &   8322.477  &    3516      & 0.574(13)  &  3 \\
\object{$\sigma$ Gem}$^d$ & 37.5$\pm$1.1 & 8346.927 &   4745      & 5.269(14)  &  0 \\
         &              &   8904.569  &    1438      & 3.43(4)    &  0 \\
         &              &   8735.159  &    7940      & 4.753(8)   &  0 \\
\object{DK Dra}   & 138$\pm$10   &   9272.484  &    4156      & 1.32(2)    & 41 \\
\object{$\epsilon$ UMi} & 106$\pm$7.6 & 8328.919 & 14690      & 0.624(6)   & 40 \\
\object{DR Dra}$^e$ & 103.$\pm$8.5&  8863.981  &   10576      & 1.554(8)   & 31 \\
\object{HR 7428}$^e$ & 323$\pm$53 &  9088.631  &   24889      & 0.112(2)   &  0 \\
\object{IM Peg}   & 96.8$\pm$7.1 &   8589.200  &    6335      & 1.692(15)  & 44 \\
         &              &   8769.630  &    8150      & 1.422(12)  & 44 \\
         &              &   8973.263  &   22143      & 1.909(4)   & 44 \\ 
\object{$\lambda$ And}$^{f, g}$ & 25.8$\pm$0.5 & 8448.155 & 31165      & 4.077(2)   &  0 \\
\end{tabular}
\end{table}
\nocite{bauebreg} \nocite{singeaa} \nocite{ortoea} \nocite{whitea}
\nocite{weltrams} \nocite{yielea} \nocite{singeab}

To compare systems at different distances, we multiply the ROSAT
countrate and the flux of the emission cores for each system with
the square of the distance listed in Table~\ref{tabros};
for \object{M$\,$67} we adopt a distance of 850\,pc (Twarog \& Anthony-Twarog 1989). 
No corrections are made for interstellar absorption.
The choice  of the 0.4--2.4\,keV bandpass minimizes the effects of interstellar
absorption, which are severe at energies $<0.4$\,keV.
As it is unknown which component of the binary emits the X-rays,
we plot the total X-ray and \ion{Ca}{ii} fluxes, adding the contributions
of both components where these are given separately by 
Fern\'andez-Figueroa et al.\ (1994).
\nocite{twan}

\begin{figure}
\centerline{\psfig{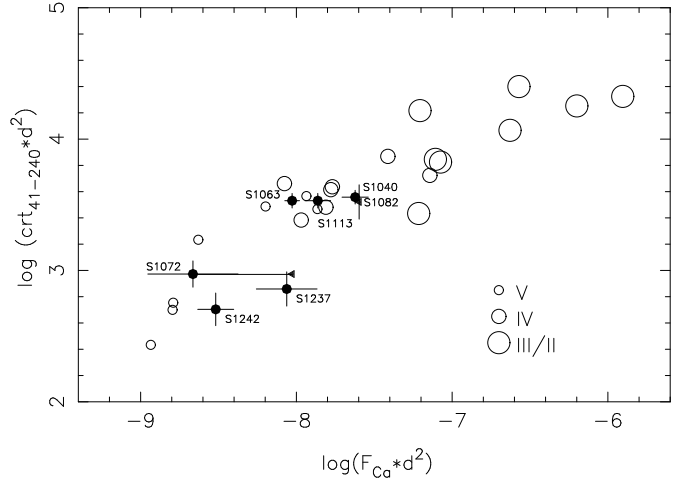} {\hfil}}
\caption{PSPC countrates in channels 41--240 versus observed \ion{Ca}{ii} H\&K 
emission flux $F_{\mathrm{Ca}}$ (in erg s$^{-1}$cm$^{-2}$). Both are 
multiplied with the square of the distance (in pc). Open circles are 
chromospherically active 
binaries from the sample of Fern\'andez-Figueroa (1994). Their size indicates
the luminosity class of the active component. When more 
than one PSPC observation 
is available, the countrate of the longest exposure is plotted.
Filled symbols are the \object{M\,67}-sources. Triangles indicate upper limits.} 
\label{xca}
\end{figure}

The resulting 'absolute' countrates and fluxes are shown in Fig.~\ref{xca}. 
The \object{M$\,$67} systems with \ion{Ca}{ii} H\&K emission clearly visible in
Fig.~\ref{cahk}, viz.\ S\,1063, S\,1113, S\,1040 and S\,1242
lie on the relation between X-ray and \ion{Ca}{ii} H\&K emission defined 
by the RS CVn systems, in agreement with the hypothesis that the
X-ray flux of these objects is related to the magnetic activity.
It is also seen that the upper limits or marginally detected
emission cores in  S\,1082, S\,1072 and S\,1237 are high enough
that we cannot exclude the hypothesis that the X-ray emission in these
systems is related to magnetic activity.

\begin{figure}
\centerline{\psfig{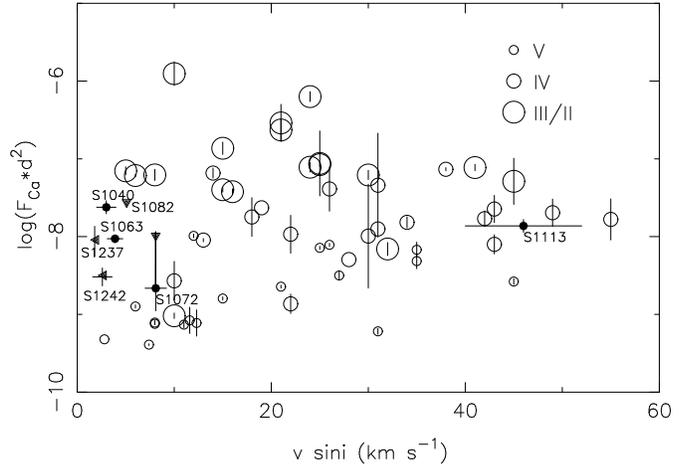} 
{\hfil}}
\caption{Observed \ion{Ca}{ii} H\&K emission flux $F_{\mathrm{Ca}}$ versus
$v\sin i$. Open circles show the comparison sample of chromospherically
active binaries up to 60 km s$^{-1}$ (Strassmeier et al. 1993 and Fern\'andez-Figueroa 1994).  
Their size indicates the luminosity class of the active
star. Vertical bars indicate 1-$\sigma$ errors due to uncertainty in the distance,
for systems with a Hipparcos parallax. \object{M\,67} sources are plotted as filled symbols. \label{vsinica}
}
\end{figure}

The rotational velocity is another indicator of magnetic activity.
We investigate the relation between rotational velocity and Ca emission
by selecting those stars from the sample of 
Fern\'andez-Figueroa for which a value of $v\sin i$ is given in 
the Catalogue of Chromospherically Active Binary Stars 
(Strassmeier et al. 1993).
In Fig.~\ref{vsinica} the \ion{Ca}{ii} H\&K emission of these stars
is compared with their $v\sin i$.
In this figure we do discriminate between the separate 
contributions of both stars to $F_{\mathrm{Ca}}$, with the 
exception of S\,1113 for which we combine the total flux
$F_{\mathrm{Ca}}$ with the $v\sin i$ of the primary.
The \object{M$\,$67} stars are found within the range occupied by 
chromospherically active stars. We note that the
correlation between the observed H\&K flux and $v\sin i$ is not tight.
In particular, high and low \ion{Ca}{ii} H\&K emission flux is found at
low values of $v\sin i$. Some of the scatter may be due to the use of 
$v\sin i$ instead of the stellar rotation period.

Parameters depending on the spectral type 
(e.g. properties of the convective region) have been used 
to reduce the scatter in the activity-rotation relation;
whereas this is successful for main-sequence stars with
$0.5\ltap B-V\ltap 0.8$, it fails for other main-sequence stars and
for giants (see discussion in St\c{e}pie\'n 1994). For example, 
the three giants \object{33 Psc} (K0 III), \object{12 Cam} (K0 III) and 
\object{DR Dra} (K0-2 III)
have $v\sin i$ values of 10, 10 and 8 km s$^{-1}$, respectively, but 
differ in $\log d^2 F_{\mathrm{Ca}}$ by three orders of magnitude (see Fig.
\ref{vsinica}).
\nocite{straea} \nocite{step}

\nocite{fernea}

\nocite{straea}

\subsection{Activity indicators}

\begin{figure}
\centerline{\psfig{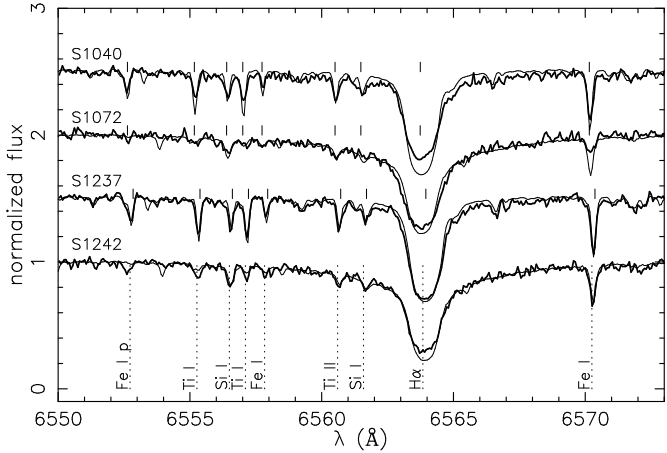} {\hfil}}
\caption{H$\alpha$ profiles of S\,1040, S\,1072, S\,1237 and S\,1242. 
Library UES-spectra (Montes \&\ Mart\'{\i}n 1998) are plotted 
with a thinner line. The comparison star is a K0III giant (\object{HD\,48432})
for S\,1040 and S\,1237, and a G0IV--V subgiant (\object{HD\,160269})
for the other stars. Position of H$\alpha$ and other lines are indicated
with vertical lines.}
\label{halpha}
\end{figure}

The H$\alpha$ lines ($\lambda$ 6562.76 \AA) of only two stars, S\,1063 and S\,1113, show
clear evidence of emission, as shown in Fig.~\ref{subha}. This is described
in more detail in their individual subsection in Sect.~4.3. 
For the other \object{M$\,$67} stars, we have used the few (sub)giants in the library of UES-spectra 
of Montes \&\ Mart\'{\i}n (1998) to investigate the behaviour of H$\alpha$. 
We have chosen library spectra of stars that match the spectra of the 
\object{M$\,$67} stars as closely as 
possible (see Fig.~\ref{halpha}). 
S\,1072 and S\,1237 show no evidence for filling in of the H$\alpha$ profile compared 
to a G0IV--V and a K0III star, respectively. In S\,1040 H$\alpha$ seems slightly filled in compared to 
a G8IV and a K0III star. This is also the case for S\,1242 compared to a G0IV-V star, but we note 
that no classification for this star is found in literature. For S\,1082, no matching spectrum is 
available in this wavelength region.
\nocite{montmart}

Filling in of the lines in the \ion{Mg}{i} b triplet ($\lambda\lambda$ 
5167.33, 5172.70 and 5183.62 \AA) and in the \ion{Na}{i} D doublet 
($\lambda\lambda$ 5889.95 and 5895.92 \AA) is visible in some active stars. 
The presence of a \ion{He}{i} D$_3$ ($\lambda$ 5876.56) absorption or 
emission feature can also indicate activity (see discussion in 
Montes \&\ Mart\'{\i}n (1998) and references therein). However, in none of the 
\object{M$\,$67} stars we see filled in \ion{Mg}{i} b and \ion{Na}{i} D lines. 
Neither do we see a clear \ion{He}{i} D$_3$ feature.
For S\,1082 (\ion{Mg}{i} b and \ion{Na}{i} D) and S\,1113 (\ion{Mg}{i} b,
 \ion{Na}{i} D and  \ion{He}{i} D$_3$) we find no suitable library stars for 
these features. 

\subsection{Individual systems}

\subsubsection{S\,1063 and S\,1113}

\begin{figure}
\centerline{\psfig{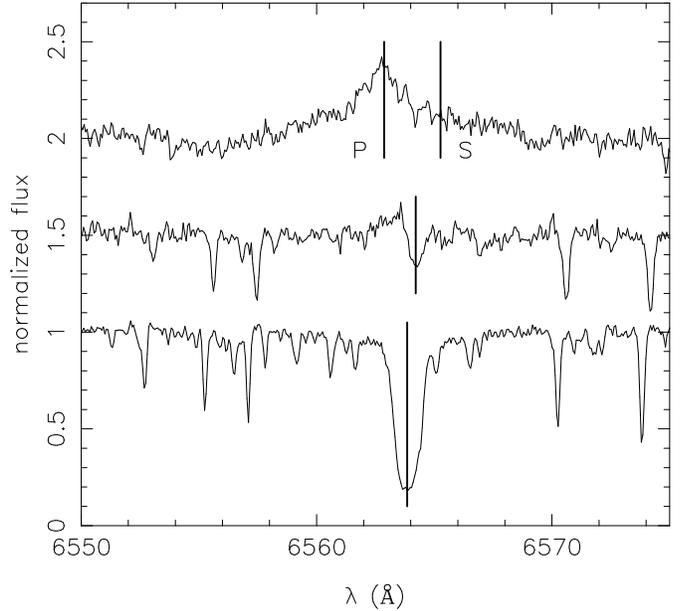}}
\caption{H$\alpha$ in S\,1113 (top), S\,1063 (middle) and comparison giant 
S\,1288. The intensity is normalized to the continuum level; the upper two 
spectra are displaced vertically by 0.5 and 1 unit. The line marks the radial 
velocity shift of the stars as determined by the crosscorrelation; for S\,1113 the 
primary (P) and secondary (S) star are separately indicated.} 
\label{subha}
\end{figure}

The two stars below the subgiant branch, S\,1063 and S\,1113,
both show relatively strong \ion{Ca}{ii} H\&K emission, and are the
only two stars in our sample  showing  H$\alpha$ in emission, shown in
Fig.~\ref{subha}.
We use the orbital solutions for both objects to try and identify
the star responsible for these emission lines. The velocities of
both components in S\,1113, and of one component in S\,1063
are indicated in Fig.~\ref{subha}.

The H$\alpha$ line profile of S\,1063 is asymmetric, showing 
emission which is blue-shifted with respect to the absorption.
The location of the absorption line is compatible with the velocity
of the primary, which dominates the flux; the emission is probably due
to the secondary. 
Remarkably, the \ion{Ca}{ii} H\&K emission peak is at the velocity of the
primary. This suggests that the H$\alpha$ emission is not chromospheric
in nature.
The H$\alpha$ emission of S\,1063 does not show the double peak that is
known to indicate accretion disk emission (Horne \&\ Marsh, 1986).
\nocite{hornmars}

\begin{figure}
\centerline{\psfig{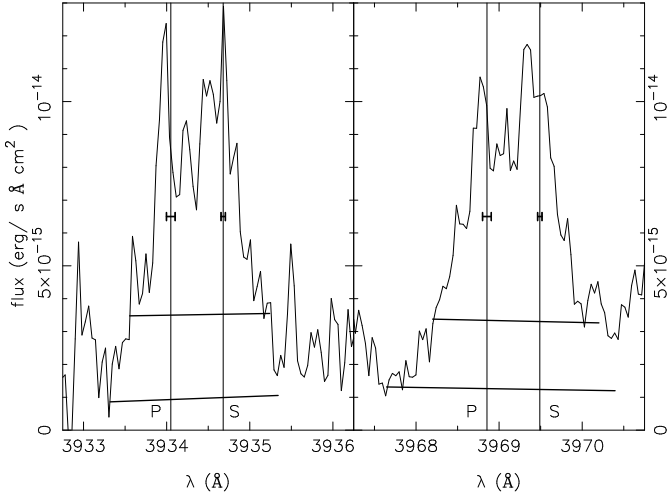}}
\caption{\ion{Ca}{ii} H\&K emission cores in S\,1113. The vertical lines
mark the shifted position of the lines for the primary (P) and secondary (S)
as derived from the crosscorrelation with the radial velocity standard 
\object{HD$\,$132737} 
(K0III, ESA 1997). Errors are also indicated. The long horizontal lines indicate the upper and lower limit chosen
to determine the emitted \ion{Ca}{ii} H\&K flux (see Sect.~3.1).}
\label{subca}
\end{figure}

In S\,1113 the H$\alpha$ emission profile is symmetric and broad,
with full width at continuum level of 15 \AA.                
The emission peak is centered on the more massive star, 
which contributes 82\%\ of the total flux (Table~\ref{fluxes}). 
This suggests that the H$\alpha$ emission is due to the primary.
The \ion{Ca}{ii} H\&K emission shows marginal evidence for a double
peak, suggesting that both stars contribute to the chromospheric
emission. In Fig.~\ref{subca} we indicate the expected position of 
the H\&K lines for both stars. For the phase observed, their peaks overlap 
in the crosscorrelation function. In the figure we use the velocities 
resulting from fitting the order that gives the 'cleanest' crosscorrelation.

\subsubsection{S\,1072 and S\,1237}

The \ion{Ca}{ii} H\&K emission in the wide binaries S\,1072 and S\,1237
is only marginally significant. The level of their X-ray and \ion{Ca}{ii} 
emission is more appropriate for active main-sequence stars (Fig.\ref{xca}).
One might speculate that it is due to the invisible companion of the 
giant detected in the crosscorrelation; even if this were the case,
we would not understand why this companion would be chromospherically active.
We conclude that we do not understand why these two stars are X-ray
sources. We find no indication for a faint secondary in the crosscorrelation
profile of S\,1237; at the time of observation, the spectra of two equally 
massive stars as suggested by Janes \&\ Smith (1984) would be separated by 
4 to 7 km s$^{-1}$ (derived from the ephemeris in Mathieu et al. 1990).
Since the secondary is 1.6 magnitude fainter in V than the primary, we think that
this small separation is compatible with finding a single peak in the 
crosscorrelation.

\subsubsection{S\,1242}

S\,1242 is chromospherically active, as shown by its 
\ion{Ca}{ii} H\&K emission. 
We suggest that this activity, which also explains the X-rays,
is due to rapid rotation induced by tidal interaction at periastron, which
tries to bring the subgiant into corotation with the orbit at
periastron.
If we assume that the observed period of photometric variability
is the rotation period we derive an inclination of 
$\sim$ 9$\degr$ using our maximum value of $v\sin i$ and the estimated 
radius. This would be in agreement with a companion
at the high end of the range 0.14--0.94 $M_{\odot}$ allowed by the mass 
function (Mathieu et al. 1990).

\subsubsection{S\,1040}

Our detection of clear chromospheric emission indicates that
the X-ray emission of S\,1040 is due to the giant.
The white dwarf has a low temperature, and is unlikely to contribute
to the X-ray flux.
We find a rather slow rotational velocity for the giant, 3\,km s$^{-1}$.
Gilliland et al.\ (1991) detected a periodicity of 7.97 days in
the visual flux  (B and V bandpasses) of S\,1040, with an
amplitude of 0.012\,mag.
If this is the rotation period of the giant, the radius of 5.1\,$R_{\sun}$
(Landsman et al.\ 1997) implies an equatorial rotation
velocity of $v=32$\,km s$^{-1}$.
This is compatible with the velocity measured with our crosscorrelation,
$v\sin i$, for an inclination $i\ltap5.3\degr$.
This inclination has an a priori probability less than 0.5\%; and it
implies an unacceptably high mass for the white dwarf, from the
measured mass function $f(m)=0.00268$.
We conclude that the 8\,days period cannot be the rotation period
of the giant.
It is doubtful that the white dwarf can be responsible, as its
contribution to the $B$ and $V$ flux is small.

\subsubsection{S\,1082}

\begin{figure}
\centerline{\psfig{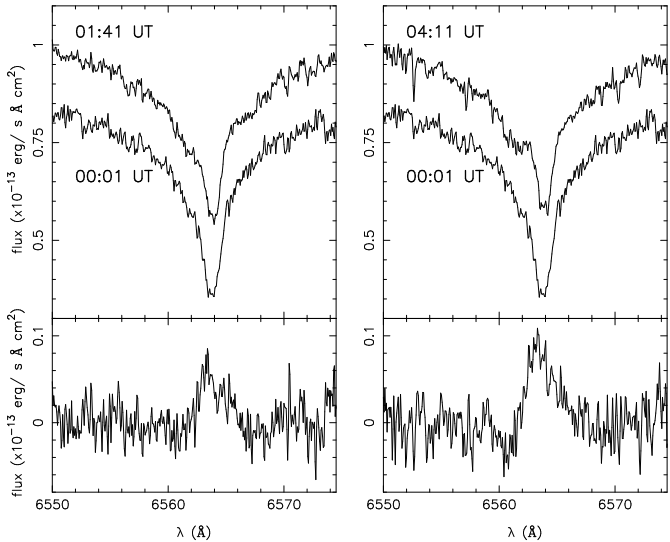}}
\caption{H$\alpha$ profile for the three exposures of S\,1082.  
The spectra are labelled according to the UT at start of the exposure.
For clarity, the lower spectrum is offset with $-0.2 \times 10^{-13}$
erg s$^{-1}$ \AA$^{-1}$ cm$^{-2}$ in both figures. The lower panels show
the difference between the first and second (left) and the first and the 
third exposure.} 
\label{s1082ha}
\end{figure}

\begin{figure}
\centerline{\psfig{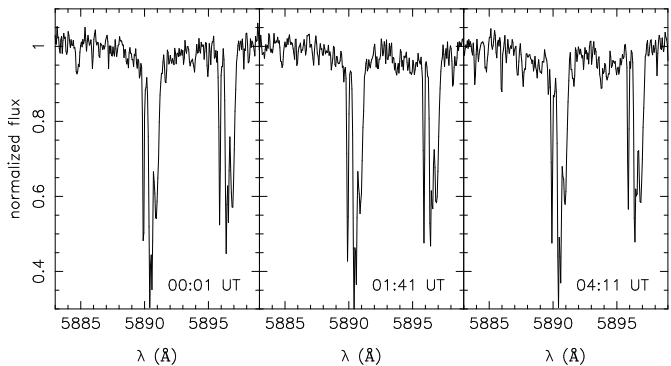}}
\caption{\ion{Na}{i} D lines in S\,1082. Shown from left to right are the
three spectra labelled according to the UT at start of the exposure. 
Variability is most clearly seen left of the Na $\lambda$ 5895.92 line.}
\label{s1082na}
\end{figure}

The H$\alpha$ absorption profile of the blue straggler S\,1082
is variable.
If we consider the most symmetric spectrum profile, that of 
00:01 UT, as the unperturbed profile of the primary, we find
that the changes are due to extra emission.
This is illustrated in Fig.~\ref{s1082ha}.
We suggest that this variation is due to the subluminous companion,
possibly to a wind of that star. We have also investigated the presence
of a broad shallow depression underlying the \ion{Na}{i} D lines (near
$\lambda$ 5895 \AA) and the \ion{O}{i} triplet (near $\lambda$ 7775 \AA)
as found by Mathys (1991). We find that this broad component is variable,
as illustrated in Fig.~\ref{s1082na}. 
Mathys (1991) suggests that the broad component originates 
in the subluminous companion.
This companion outshines the primary by a factor six at $\lambda$ 1500 
\AA\, and thus is presumably hot (Landsman et al. 1998). 
We note that the star cannot be too hot or it would not show neutral 
lines. \nocite{landea98}

\section{Discussion and conclusions}
In this paper we have tried to find an explanation for the X-ray emission
of seven sources in \object{M$\,$67}.

For S\,1242 and S\,1040 we have concluded from the \ion{Ca}{ii} H\&K 
emission cores that magnetic activity is responsible for the X-rays. This is 
supported by filling in of H$\alpha$ (see e.g. Montes et al. 1997; Eker et al. 
1995). In S\,1242, activity is likely to be triggered by interaction at
periastron in the eccentric orbit. This is also reflected in the period of 
photometric variability. For S\,1040, the reason for activity is 
less clear. The explanation could involve mass transfer from the precursor of 
the white dwarf to the giant and the latter's subsequent expansion during the 
giant phase. As was already noted by Landsman et al. (1997), a similar system 
is \object{AY Cet}, a binary of a white dwarf of $T_\mathrm{eff}=18\,000$ K 
(Simon et al. 1985) and a G5III giant in an 56.8 days circular orbit. 
The $v\sin i$ of that giant is also low, 4 km s$^{-1}$, and the long photometric period
of 77.2 days implies asynchronous rotation (Strassmeier et al. 1993). The 
X-ray luminosity for \object{AY Cet} is $1.5 \times 10^{31}$ erg s$^{-1}$ 
in the 0.2--4 keV band as measured with Einstein by Walter \&\ Bowyer (1981),
somewhat higher than the luminosity of S\,1040. 
(With the coronal model discussed by Belloni et al. (1998), the countrate 
for S\,1040 corresponds to $5.6 \times 10^{30}$ erg s$^{-1}$ in the 
0.2--4 keV band.)
Walter \&\ Bowyer attribute 
the X-rays to coronal activity of the giant.
\nocite{simoea} \nocite{waltbowy} \nocite{ekerea} \nocite{mfea}

The \ion{Ca}{ii} H\&K emission cores in S\,1063 and S\,1113 are very strong.
In S\,1113 we might even see emission 
of both stars. Due to the shape of the H$\alpha$ emission we cannot conclude 
with certainty that the X-rays arise in an active corona and not in a disk or 
stream. The wings in the emission peak of S\,1113 are very broad.
However, Montes et al. (1997) have demonstrated that the excess emission in 
the H$\alpha$ lines of the more active binaries is sometimes a composite of a 
narrow and broad component, the latter having a full width at half maximum of 
up to 470 km s$^{-1}$. They ascribe this broad component to microflaring accompanied 
by large scale motions. We note the similarity between S\,1113 and 
\object{V711 Tau}, 
a well known extremely active binary of a G5IV ($v\sin i$ = 13 km s$^{-1}$) and K1IV 
($v\sin i = 38$ km s$^{-1}$) star in a 2.84 days circular orbit and a mass ratio 0.79 
(Strassmeier et al. 1993); the mass ratio of S\,1113 is 0.70 
(Mathieu et al. 1998, in preparation). From the countrate of 
\object{V711 Tau} in Table~\ref{tabros}
we find $L_{\mathrm{x}}=6.8 \times 10^{30}$ erg s$^{-1}$ in the 0.1--2.4 keV band using
the same model as in Belloni et al. (1998) with $N_{H}=0$, which is comparable 
to the luminosity of S\,1113 in the same band $L_{\mathrm{x}}=7.3 \times 10^{30}$ 
erg s$^{-1}$ (Belloni et al. 1998). The H$\alpha$ emission of S\,1063 is more 
difficult to explain. As this system is not double-lined, H$\alpha$ emisson by
the (invisible) secondary star would have to be strong to rise above the 
continuum of the primary.
\nocite{mfea}

In the binaries S\,1072 and S\,1237 we see no H$\alpha$ emission while
the level of \ion{Ca}{ii} H\&K emission is low in comparison with active
stars of the same luminosity class. We have no explanation for this. For 
S\,1072, an option is a wrong identification of the X-ray source with an 
optical counterpart. Belloni et al. (1998) give a probability of 43\% 
that one or two of their twelve identifications of an X-ray source with a 
binary in \object{M$\,$67} is due to chance.

No \ion{Ca}{ii} H\&K emission is seen in the spectrum of the blue straggler
S\,1082. Possibly, the X-ray emission has to do with the hot, subluminous
secondary that could also cause the photometric
variability and whose signature we might have seen in the H$\alpha$ line.

\begin{acknowledgements}
The authors wish to thank G. Geertsema for her help during our
observations, P. Groot for providing his program to compute projected 
rotational velocities with the Fourier-Bessel transformation method and 
M. van Kerkwijk for comments on the manuscript. MvdB is supported by
the Netherlands Organization for Scientific Research (NWO).
\end{acknowledgements}



\begin{thebibliography}{}

\bibitem[\protect\astroncite{Allen \& Strom}{1995}]{allestro}
Allen, L.~E., Strom, K.~M. 1995, AJ, 109, 1379

\bibitem[\protect\astroncite{Bauer \& Bregman}{1996}]{bauebreg}
Bauer, F., Bregman, J.~N. 1996, ApJ, 457, 382

\bibitem[\protect\astroncite{Belloni et~al.}{1998}]{bellea}
Belloni, T., Verbunt, F., Mathieu, R.~D. 1998, A\&A, 339, 431

\bibitem[\protect\astroncite{Belloni et~al.}{1993}]{bellea93}
Belloni, T., Verbunt, F., Schmitt, J. H. M.~M. 1993, A\&A, 269, 175

\bibitem[\protect\astroncite{Carter et~al.}{1993}]{cartea}
Carter, D., Benn, C.~R., Rutten, R. G.~M., Breare, J.~M., Rudd, P.~J., King,
  D.~L., Clegg, R. E.~S., Dhillon, V.~S., Arribas, S., Rasilla, J.-L., Garcia,
  A., Jenkins, C.~R., Charles, P.~A. 1993,
\newblock ISIS Users' Manual,
\newblock http://www.ing.iac.es

\bibitem[\protect\astroncite{Eker et~al.}{1995}]{ekerea}
Eker, Z., Hall, D.~S., Anderson, C.~M. 1995, ApJS, 96, 581

\bibitem[\protect\astroncite{ESA}{1997}]{hipp}
ESA 1997,
\newblock The Hipparcos and Tycho Catalogues, ESA SP 1200

\bibitem[\protect\astroncite{Fern\'andez-Figueroa et~al.}{1994}]{fernea}
Fern\'andez-Figueroa, M.~J., Montes, D., Castro, E.~D., Cornide, M. 1994, ApJS,
  90, 433

\bibitem[\protect\astroncite{Gilliland et~al.}{1991}]{gillea}
Gilliland, R.~L., Brown, T.~M., Duncan, D.~K., Suntzeff, N.~B., Lockwood,
  G.~W., Thompson, D.~T., Schild, R.~E., Jeffrey, W.~A., Penrase, B.~E. 1991,
  AJ, 101, 541

\bibitem[\protect\astroncite{Goranskij et~al.}{1992}]{goraea}
Goranskij, V.~P., Kusakin, A.~V., Mironov, A.~V., Moshkaljov, V.~G.,
  Pastukhova, E.~N. 1992, Astron. Astrophys. Trans., 2, 201

\bibitem[\protect\astroncite{Groot et~al.}{1996}]{grootea}
Groot, P.~J., Piters, A. J.~M., van Paradijs, J. 1996, A\&AS, 118, 545

\bibitem[\protect\astroncite{Gunn et~al.}{1996}]{gunnea}
Gunn, A.~G., Hall, J.~C., Lockwood, G.~W., Doyle, J.~G. 1996, A\&A, 305, 146

\bibitem[\protect\astroncite{Hobbs \& Mathieu}{1991}]{hobbmath}
Hobbs, L.~M., Mathieu, R.~D. 1991, PASP, 103, 431

\bibitem[\protect\astroncite{Horne}{1986}]{horn}
Horne, K. 1986, PASP, 98, 609

\bibitem[\protect\astroncite{Horne \& Marsh}{1986}]{hornmars}
Horne, K., Marsh, T.~R. 1986, MNRAS, 218, 761

\bibitem[\protect\astroncite{Janes \& Smith}{1984}]{janesmit}
Janes, K.~A., Smith, G.~H. 1984, AJ, 89, 487

\bibitem[\protect\astroncite{Kaluzny \& Radczynska}{1991}]{kara}
Kaluzny, J., Radczynska, J. 1991,
\newblock IBVS 3586

\bibitem[\protect\astroncite{King}{1985}]{king}
King, D.~L. 1985, ING Technical Note, 31

\bibitem[\protect\astroncite{Kippenhahn \& Weigert}{1967}]{kippweig}
Kippenhahn, R., Weigert, A. 1967, Zeitschrift f\"ur Astrophysik, 65, 251

\bibitem[\protect\astroncite{Kurochkin}{1960}]{kuro}
Kurochkin, N.~E. 1960, Astron. Circular USSR, 212, 9

\bibitem[\protect\astroncite{Landsman et~al.}{1997}]{landea}
Landsman, W., Aparicio, J., Bergeron, P., {Di Stefano}, R., Stecher, T.~P.
  1997, ApJ, 481, L93

\bibitem[\protect\astroncite{Landsman et~al.}{1998}]{landea98}
Landsman, W., Bohlin, R.~C., Neff, S.~G., O'Conell, R.~W., Roberts, M.~S.,
  Smith, A.~M., Stecher, T.~P. 1998, AJ, 116, 789

\bibitem[\protect\astroncite{Latham et~al.}{1992}]{lathmathea}
Latham, D.~W., Mathieu, R.~D., Milone, A. A.~E., Davis, R.~J. 1992,
\newblock in A. Duquennoy, M. Mayor (eds.), Binaries as tracers of stellar
  formation, Cambridge University Press, Cambridge,  132

\bibitem[\protect\astroncite{Massey et~al.}{1988}]{massea}
Massey, P., Strobel, K., Barnes, J.~V., Anderson, E. 1988, ApJ, 328, 315

\bibitem[\protect\astroncite{Mathieu \& Latham}{1986}]{mathlath}
Mathieu, R.~D., Latham, D.~W. 1986, AJ, 92, 1364

\bibitem[\protect\astroncite{Mathieu et~al.}{1990}]{mathlathea}
Mathieu, R.~D., Latham, D.~W., Griffin, R.~F. 1990, AJ, 100, 1859

\bibitem[\protect\astroncite{Mathieu et~al.}{1986}]{mathea86}
Mathieu, R.~D., Latham, D.~W., Griffin, R.~F., Gunn, J.~E. 1986, AJ, 92, 1100

\bibitem[\protect\astroncite{Mathys}{1991}]{math}
Mathys, G. 1991, A\&A, 245, 467

\bibitem[\protect\astroncite{Montes et~al.}{1997}]{mfea}
Montes, D., Fern\'andez-Figueroa, M.~J., {De Castro}, E., Sanz-Forcada, J.
  1997, A\&AS, 125, 263

\bibitem[\protect\astroncite{Montes \& Mart\'{\i}n}{1998}]{montmart}
Montes, D., Mart\'{\i}n, E.~L. 1998, A\&AS, 128, 485

\bibitem[\protect\astroncite{Montgomery et~al.}{1993}]{montea}
Montgomery, K.~A., Marshall, L.~A., Janes, K.~A. 1993, AJ, 106, 181

 11 \bibitem[\protect\astroncite{Nissen et~al.}{1987}]{nissea}
Nissen, P.~E., Twarog, B.~A., Crawford, D.~L. 1987, AJ, 93, 634

\bibitem[\protect\astroncite{Ortolani et~al.}{1997}]{ortoea}
Ortolani, A., Maggio, A., Pallavicini, R., Sciortino, S., Drake, J.~J., ADrake,
  S. 1997, A\&A, 325, 664

\bibitem[\protect\astroncite{Pasquini \& Belloni}{1998}]{pasqbell}
Pasquini, L., Belloni, T. 1998, A\&A, 336, 902

\bibitem[\protect\astroncite{Piters et~al.}{1996}]{pitegrooea}
Piters, A. J.~M., Groot, P.~J., van Paradijs, J. 1996, A\&AS, 118, 529

\bibitem[\protect\astroncite{Pritchet \& Glaspey}{1991}]{pritglas}
Pritchet, C.~J., Glaspey, J.~W. 1991, ApJ, 373, 105

\bibitem[\protect\astroncite{Rajamohan et~al.}{1988}]{rajaea}
Rajamohan, R., Bhattacharyya, J.~C., Subramanian, V., Kuppuswamy, K. 1988,
  Bull. Astr. Soc. India, 16, 139

\bibitem[\protect\astroncite{Sanders}{1977}]{san}
Sanders, W.~L. 1977, A\&AS, 27, 89

\bibitem[\protect\astroncite{Simoda}{1991}]{simo}
Simoda, M. 1991,
\newblock IBVS 3675

\bibitem[\protect\astroncite{Simon et~al.}{1985}]{simoea}
Simon, T., Fekel, F.~C., Gibson, D.~M. 1985, ApJ, 295, 153

\bibitem[\protect\astroncite{Singh et~al.}{1996a}]{singeaa}
Singh, K.~P., Drake, S.~A., White, N.~E. 1996a, AJ, 111, 2415

\bibitem[\protect\astroncite{Singh et~al.}{1996b}]{singeab}
Singh, K.~P., Drake, S.~A., White, N.~E. 1996b, AJ, 112, 221

\bibitem[\protect\astroncite{St\c{e}pie\'n}{1994}]{step}
St\c{e}pie\'n 1994, A\&A, 292, 191

\bibitem[\protect\astroncite{Strassmeier et~al.}{1993}]{straea}
Strassmeier, K.~G., Hall, D.~S., Fekel, F.~C., Scheck, M. 1993, A\&AS, 100, 173

\bibitem[\protect\astroncite{Tonry \& Davis}{1979}]{tonrdavi}
Tonry, J., Davis, M. 1979, AJ, 84, 1511

\bibitem[\protect\astroncite{Twarog \& Anthony-Twarog}{1989}]{twan}
Twarog, B.~A., Anthony-Twarog, B.~J. 1989, AJ, 97, 759

\bibitem[\protect\astroncite{Unger et~al.}{1993}]{ungeea}
Unger, S., Walton, N., Pettini, M., Tinbergen, J. 1993,
\newblock UES Users' Manual,
\newblock http://www.ing.iac.es/

\bibitem[\protect\astroncite{Verbunt \& Phinney}{1995}]{verbphin}
Verbunt, F., Phinney, E.~S. 1995, A\&A, 296, 709

\bibitem[\protect\astroncite{Walter \& Bowyer}{1981}]{waltbowy}
Walter, F.~M., Bowyer, S. 1981, ApJ, 245, 671

\bibitem[\protect\astroncite{Welty \& Ramsey}{1995}]{weltrams}
Welty, A.~D., Ramsey, L.~W. 1995, AJ, 109, 2187

\bibitem[\protect\astroncite{White et~al.}{1994}]{whitea}
White, N.~E., Arnaud, K., Day, C. S.~R., Ebisawa, K., Gotthelf, E.~V., Mukai,
  K., Soong, Y., Yaqoob, T., Antunes, A. 1994, PASJ, 46, L97

\bibitem[\protect\astroncite{Wyatt}{1985}]{wyat}
Wyatt, W.~F. 1985,
\newblock in A.~G. {Davis Philip}, D.~W. Latham (eds.), Stellar radial
  velocities, Vol.~88 of {\em IAU Coll.\/}, L. Davis Press, Inc., Schenectady,
  N.Y.,  123

\bibitem[\protect\astroncite{Yi et~al.}{1997}]{yielea}
Yi, Z., Elgar{\o}y, {\O}., Engvold, O., Westergaard, N.~J. 1997, A\&A, 318, 791

\bibitem[\protect\astroncite{Zhilinskii \& Frolov}{1994}]{zhilfrol}
Zhilinskii, E.~G., Frolov, V.~N. 1994, Astronomy Letters, 20, 80

\bibitem[\protect\astroncite{Zimmermann et~al.}{1994}]{zimmea94}
Zimmermann, H., Becker, W., Belloni, T., D\"obereiner, S., Izzo, C., Kahabka,
  P., Schwentker, O. 1994,
\newblock EXSAS User's Guide: Extended scientific analysis system to evaluate
  data from the astronomical X-ray satellite ROSAT, Technical Report 257,
\newblock MPE

\end{thebibliography}
\end{document}